\documentstyle[12pt,epsf]{article}

0

\begin{document}

\newcommand{\MqV}{$\cal M$$_{q}$($\cal V$)}
\def\II{\relax{\rm 1\kern-.35em1}}
\def\IP{\relax{\rm I\kern-.18em P}}
\renewcommand{\theequation}{\thesection.\arabic{equation}}
\csname @addtoreset\endcsname{equation}{section}

\begin{flushright}
hep-th/9512017

IMAFF/95-39
\end{flushright}
\vglue 1cm

\begin{center}

{}~\vfill

{\large \bf $S$-Duality and the Calabi-Yau Interpretation}
\vspace{10 mm}
{\large \bf of the $N\!=\!4$ to $N\!=\!2$ Flow.}

\vspace{15 mm}

\end{center}

\begin{center}

{\bf C{\'e}sar G{\'o}mez$^{a}$, Rafael Hern{\'a}ndez$^{b}$ and
Esperanza L{\'o}pez$^{a}$}\footnote{Address after January 1996, Institute
for Theoretical
Physics, University of California, Santa Barbara, CA 93106-4030.}

\vspace{8 mm}
$^{a}${\em Instituto de Matem{\'a}ticas y F{\'\i}sica Fundamental, CSIC
\protect \\
Serrano 123, 28006 Madrid, Spain}
\vspace{8mm}

$^{b}${\em Departamento de F{\'\i}sica Te{\'o}rica, C-XI, Universidad
Aut{\'o}noma
de Madrid \protect \\ Cantoblanco, 28049 Madrid, Spain}

\end{center}

\vspace{15mm}

\begin{abstract}
The action of the $S$-duality $Sl(2,Z)$ group on the moduli of
the Calabi-Yau manifold $W\IP^{12}_{11226}$ appearing in the
rank two dual pair $(K^{3}\times T^{2}/W\IP^{12}_{11226})$ is
defined by interpreting the $N\!=\!4$ to $N\!=\!2$ flow, for
$SU(2)$ supersymmetric Yang-Mills, in terms of the Calabi-Yau
moduli. The different singularity loci are mapped in a one to
one way, and the ($N\!=\!2$ limit/point particle limit) is
obtained in both cases by the same type of blow up. Moreover, it
is shown that the $S$-duality group permutes the different
singularity loci of the moduli of $W\IP^{12}_{11226}$. We study
the transformation under $S$-duality of the Calabi-Yau Yukawa
couplings.

\end{abstract}

\vspace{2cm}

\pagebreak


\section{Introduction.}

The discovery of heterotic-type II dual pairs \cite{r1}-\cite{KV}
opens the possibility to enter into the realm of string
non perturbative effects. The first direct application of
heterotic-type II dual pairs will be its use to derive the
exact quantum moduli space of quantum field theories defined as the
low energy limit of heterotic string compactifications. This has been
done for the dual pair defined by $(K^{3}\times T^{2}/W\IP
^{12}_{11226})$ \cite{KV} by analizing the point
particle limit \cite{GL}-\cite{KKLMV} of the moduli of complex
structures of $W\IP^{12}_{11226}$, reproducing in \cite{KKLMV}
the exact results of the Seiberg-Witten solution for pure
$N\!=\!2$ $SU(2)$ supersymmetric Yang-Mills \cite{SW}.
In this letter, we will proceed in a somewhat opposite way:
instead of seeking the field theory point particle limit of
the string, we will try to read off the string theory directly from
the field theory. To do so, we will focus on pure $N\!=\!2$ $SU(2)$
supersymmetric Yang-Mills
theory from the point of view of the $N\!=\!4$ to $N\!=\!2$ flow
\cite{SW2}, i.e., we will work with the $N\!=\!2$ theory
possesing $N\!=\!4$ matter content. In this way, we start with an
{\em extended} moduli space, parametrized by $\hat{u}$
($\hat{u}\equiv \frac {u}{f}$, with $u$ the
Seiberg-Witten moduli variable, and $f=\frac {1}{4}m^{2}$ the soft breaking mass term) and $\tau$ (the $N\!=\!4$ moduli).
It will be this space the one we will put in correspondence with the
moduli of complex structures of $W\IP^{12}_{11226}$ \cite{can,CYY},
mapping, in a one to one way, the different
singular loci of the two spaces. In both cases, the pure $N\!=\!2$
theory can be
derived as the blow up of a weak coupling ($\tau \rightarrow
\infty/S \rightarrow \infty$) singular point.

The $Sl(2,Z)$ $S$-duality group, acting on the $N\!=\!4$ moduli
$\tau$, induces duality transformations on a double covering of
the moduli of complex
structures of $W\IP_{11226}^{12}$.
We study the transformations of the Yukawa couplings with
respect to $S$-duality.


\section{$N\!=\!4$ to $N\!=\!2$ Flow and Duality.}

Let us consider $N\!=\!2$ $SU(2)$ supersymmetric Yang-Mills with
one hypermultiplet in the adjoint representation. The curve
describing this model is given by \cite{SW2}\footnote{The
Weierstrass invariants $e_{i}$ can be defined in the terms of
Jacobi theta functions:
\[e_{1}=\frac {1}{3}(\theta_{2}^{4}(0,\tau)+\theta_{3}^{4}(0,\tau)),
e_{2}=- \frac {1}{3}(\theta_{1}^{4}(0,\tau)+\theta_{3}^{4}(0,\tau)),
e_{3}=\frac {1}{3}(\theta_{1}^{4}(0,\tau)-\theta_{2}^{4}(0,\tau)).\]}:
\begin{equation}
y^{2}=(x-e_{1}(\tau)\tilde{u}-e_{1}^{2}(\tau)f)
        (x-e_{2}(\tau)\tilde{u}-e_{2}^{2}(\tau)f)
        (x-e_{3}(\tau)\tilde{u}-e_{3}^{2}(\tau)f),
\label{eq:cur}
\end{equation}
where $f=\frac {1}{4}m^{2}$, with m the mass of the
hypermultiplet. In the massless limit we recover the $N\!=\!4$
curve of the elliptic moduli $\tau$. After the finite
renormalization \(u=\tilde{u} + \frac {1}{2}e_{1}(\tau)f\) \cite{SW2},
and the {\em double scaling} limit defined by
\begin{equation}
\lim_{\begin{array}{c}
        \tau \rightarrow  i \infty \\
        m \rightarrow \infty \end{array}}  2 q^{1/2}m^{2}=\Lambda^{2},
\label{eq:DSL}
\end{equation}
where \(q \equiv e^{2 \pi i \tau}\), we obtain, from
(\ref{eq:cur}), the Seiberg-Witten solution for pure $N\!=\!2$ $SU(2)$
supersymmetric Yang-Mills with $u= \langle Tr \phi^{2} \rangle$. By an
affine transformation we can put the curve (\ref{eq:cur}) in the form
\begin{equation}
y^{2}=x(x-1)(x-\lambda(\tilde{u}, \tau,f)),
\label{eq:lecur}
\end{equation}
where
\begin{equation}
\lambda(\tilde{u},\tau,f)=
\frac {(e_{3}-e_{1})(\tilde{u}+ f(e_{3}+e_{1}))}
         {(e_{2}-e_{1})(\tilde{u}+ f(e_{2}+e_{1}))}.
\label{eq:lamdef}
\end{equation}

The elliptic curve (\ref{eq:lecur}) is characterized by the
$j$-invariant
\begin{equation}
j(\lambda)= 2^8 \frac
{(\lambda^{2}-\lambda+1)^{3}}{\lambda^{2}(\lambda -1)^{2}},
\end{equation}
which satisfies the relations $j(\lambda)=j(\frac {1}{\lambda})=
j(1-\lambda)$.

Using the transformation laws of the $e_{i}(\tau)$ as modular
forms of weight two with respect to $\Gamma_{2}$, we easily
get the following set of duality relations:
\begin{equation}
1-\lambda(\tilde{u},\tau,f)=\lambda(\tilde{u}^{M},
\frac {-1}{\tau},f) \equiv \lambda(\tilde{u}',\tau,f),
\label{eq:l1}
\end{equation}
\begin{equation}
\frac{1}{\lambda(\tilde{u},\tau,f)}=\lambda
(\tilde{u},\tau +1,f) \equiv \lambda(\tilde{u}'',\tau,f)
\label{eq:l2}
\end{equation}
where we have defined
\begin{equation}
\begin{array}{l}\tilde{u}^{M} =  \tau^{2} \tilde{u}, \hspace{2.5cm}
\tilde{u}' =  {\displaystyle \frac
{a-f_{1}(\tilde{u})b}{(e_{2}-e_{1})f_{1}(\tilde{u})-
        (e_{3}-e_{1})}} \\
        \tilde{u}''  =  {\displaystyle \frac
{a-f_{2}(\tilde{u})b}{(e_{2}-e_{1})f_{2}(\tilde{u})-(e_{3}-e_{1})},}
\label{eq:utr}
\end{array}
\end{equation}
with
\begin{equation}
\begin{array}{ll}
a =  f(e_{3}^{2}-e_{1}^{2}), \hspace{2cm} &
b =  f(e_{2}^{2}-e_{1}^{2}) \\
f_{1}(\tilde{u}) =  {\displaystyle \frac {1}{\lambda(\tilde{u},
\tau,f)},} & f_{2}(\tilde{u})  =  {\displaystyle 1-
\lambda(\tilde{u},\tau,f).}
\end{array}
\end{equation}
Notice from the transformation rule for $\tilde{u}$ in (\ref{eq:utr})
that $\tilde{u}$ transforms as a modular form of weight two.

In the double scaling limit defined by (\ref{eq:DSL}) we get
\begin{equation}
\lim_{\begin{array}{c}\tau \rightarrow i \infty \\
                        m^{2} \rightarrow \infty
      \end{array}}   \lambda (\tilde{u},\tau,f)=
      \frac {u + \Lambda^{2}}{u- \Lambda^{2}} \equiv \lambda_{SW}(u),
\label{eq:lamsw}
\end{equation}
with $\lambda_{SW}(u)$ the value of $\lambda$ obtained from the
Seiberg-Witten curve
$y^{2}=(x+\Lambda^{2})(x-\Lambda^{2})(x-u)$ for pure $N\!=\!2$
supersymmetric Yang-Mills. Moreover, in this limit we notice,
using relations (\ref{eq:l1}) and (\ref{eq:l2}), that the $Sl(2,Z)$
duality transformations
$T: \tau \rightarrow \tau+1$ and $S:\tau \rightarrow \frac {-1}{\tau}$
induce, on the Seiberg-Witten plane, the transformations
$u \rightarrow -u$ and
$u \rightarrow \frac {3\Lambda^{4}+ \Lambda^{2}u}{\Lambda^{2}-u}$,
respectively.
These transformations, which generate the group
\(\Gamma_{W}\equiv Sl(2,Z)/\Gamma_{2}\)
\cite{GL2}, interchange the singularities of the pure $N\!=\!2$
theory. In other words, we observe how the $Sl(2,Z)$ duality
transformations on the $N\!=\!4$
moduli $\tau$ permute, in the double scaling limit, the phases
of pure $N\!=\!2$ theory \cite{DW}.

Next, let us study the singularity locus for the curve (\ref{eq:cur}).
We will work in the ($\hat{u}, \tau$)-plane, with
$\hat{u}\equiv \frac {u}{f}$; the use of this dimensionless
variable will prove important later on, in the correspondence
with the Calabi-Yau moduli space. The following set of regions can then
be differentiated (see Figure 1):
\begin{equation}
        \begin{array}{lcl}
        {\cal C}_{\infty} & \equiv & \{ \tau= i \infty \}, \\
        {\cal C}_{0}      & \equiv & \{ \hat{u}(\tau)=
\frac {3}{2}e_{1}(\tau) \}, \\
        {\cal C}_{c}^{(1)} & \equiv & \{ \hat{u}(\tau)
=(e_{3}+ \frac {1}{2}e_{1})(\tau) \}, \\
        {\cal C}_{c}^{(2)} & \equiv & \{ \hat{u}(\tau)
=(e_{2}+ \frac {1}{2}e_{1})(\tau) \},  \\
        {\cal C}_{1} & \equiv & \{ {\tau = 0} \}.
        \end{array}
\label{eq:loci}
\end{equation}


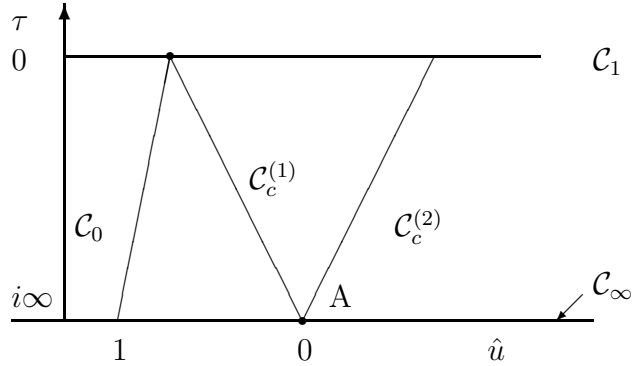
\begin{figure}[htbp]
\centering
    \begin{picture}(400,170)

    \put(100,120){\line(1,0){180}}  \put(140,120){\line(1,-2){50}}
    \put(120,20){\line(1,5){20}}  \put(240,120){\line(-1,-2){50}}

    \put(140,120){\circle*{3}}  \put(190,20){\circle*{3}}

    \put(80,130){$\tau$}  \put(80,115){$0$}  \put(80,25){$i \infty$}
    \put(118,5){$1$}  \put(188,5){$0$}  \put(200,25){A}
    \put(260,5){$\hat{u}$}  \put(300,115){${\cal C}_{1}$}
    \put(300,30){${\cal C}_{\infty}$}  \put(296,30){\vector(-1,-1){10}}
    \put(104,52){${\cal C}_{0}$} \put(225,52){${\cal C}_{c}^{(2)}$}
    \put(170,70){${\cal C}_{c}^{(1)}$}

    \thicklines
    \put(100,20){\vector(0,1){120}}  \put(80,20){\line(1,0){220}}

    \end{picture}

\caption{The different loci described in (2.11). Notice that
$e_{2}=e_{3}$ at $\tau = i \infty$, and $e_{1}=e_{3}$ at $\tau = 0$.}

\end{figure}


Notice that the loci ${\cal C}_{0}$, ${\cal C}_{c}^{(1)}$ and
${\cal C}_{c}^{(2)}$ correspond to the singularities of the curve
({\ref{eq:cur}) when written in $\hat{u}$ variables. It can be
easily proved that the duality transformations
(\ref{eq:l1}) and (\ref{eq:l2}) permute among themselves the
singular loci (\ref{eq:loci}). Namely, the transformation $S$
permutes ${\cal C}_{1}$ with ${\cal C}_{\infty}$, and ${\cal
C}_{0}$ with ${\cal C}_{c}^{(2)}$, while the locus ${\cal
C}_{c}^{(1)}$ is mapped into itself; the transformation $T$
permutes the locus ${\cal C}_{c}^{(1)}$ and ${\cal
C}_{c}^{(2)}$, and maps into itself the locus ${\cal C}_{0}$.

The loci ${\cal C}_{0}$, ${\cal C}_{c}^{(1)}$ and
${\cal C}_{c}^{(2)}$ can be described using the language
of double ramified coverings introduced in
\cite{DW} in connection with integrable models. In fact, for
the double ramified covering defined by
\begin{equation}
        \left\{ \begin{array}{lcl}
                y^{2} & = & (x-e_{1})(x-e_{2})(x-e_{3}) \\
                0     & = & t^{2}-x+ \tilde{u} ,
                \end{array}  \right.
\label{eq:ramcov}
\end{equation}
the loci ${\cal C}_{0}$, ${\cal C}_{c}^{(1)}$ and
${\cal C}_{c}^{(2)}$ are defined
by the relation $\tilde{u}(\tau)$, characterizing the values
of $\tilde{u}$ for which ({\ref{eq:ramcov}) becomes a double
{\em unramified} covering\footnote{A
massive vacuum, in the notation of reference \cite{DW}.}.


\section{Blowing up and the $N\!=\!2$ Limit.}

Let us now consider more carefully the neighbourhood of point
$A \equiv ( \hat{u}\!=\!0,\tau\!=\!i \infty)$ in the
$(\hat{u},\tau)$-plane depicted in Figure 1. Introducing a new
coordinate $\epsilon \equiv 8 q^{1/2}$, the loci
${\cal C}_{c}^{(1)}$ and ${\cal C}_{c}^{(2)}$ in the neighbourhood
of the point $A$ can be described in $(\epsilon^{2},\hat{u})$
coordinates by the parabole $\epsilon^{2}=\hat{u}^{2}$. This
parabole is tangent at the point $A$ to the locus ${\cal
C}_{\infty}=\{\tau\!=\!i \infty \}$. Using standard techniques
\cite{HARR} we
can blow up this tangency point. To do it a double blow up is
needed: in the first blow up we introduce the coordinate
$v=\frac {\epsilon^{2}}{\hat{u}}$, which transforms the tangency
into a crossing between the curve $v=\hat{u}$ and ${\cal
C}_{\infty}$; in the second step, we blow up this crossing
introducing the coordinate $w=\frac {v}{\hat{u}}$, mapping the
parabole to the line $w=1$. The exceptional divisors $E_{1}$ and
$E_{2}$, and
coordinates introduced by this blow up are described in Figure 2.


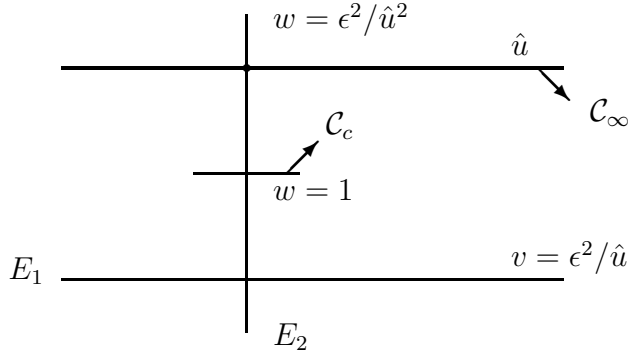
\begin{figure}[htbp]
\centering

     \begin{picture}(400,170)

     \thicklines

     \put(150,20){\line(0,1){120}}  \put(130,80){\line(1,0){40}}
     \put(80,120){\line(1,0){190}}  \put(80,40){\line(1,0){190}}

     \put(165,80){\vector(1,1){12}}  \put(260,120){\vector(1,-1){12}}
     \put(150,120){\circle*{3}}

     \put(160,135){$w=\epsilon^{2}/\hat{u}^{2}$}
     \put(280,100){${\cal C}_{\infty}$}  \put(250,125){$\hat{u}$}
     \put(180,95){${\cal C}_{c}$}  \put(250,45){$v=\epsilon^{2}/\hat{u}$}
     \put(160,70){$w=1$}
        \put(60,40){$E_{1}$} \put(160,15){$E_{2}$}

     \end{picture}

\caption{The double blow up of the tangency point A of Figure~1.}

\end{figure}


Defining $\Lambda^2=8q^{1/2}f$ \cite{SW} we observe that the
coordinate on $E_{1}$ is given by $\Lambda^4/u^2\equiv
1/\tilde{u}^2$. 

Some comments are now necessary for a proper understanding of
the physical meaning of the above construction. First of all,
the point $A$ in the $(\hat{u},\tau)$-plane we are blowing up is
a point where the value of $\lambda(\hat{u},\tau)$ is
undetermined (see equation (\ref{eq:lamdef})). In order to give
a precise meaning to the double scaling limit in the
$(\hat{u},\tau)$-plane, we need to blow up the point $A$, and to
define $\lambda_{SW}$ (see equation (\ref{eq:lamsw})) as a
function on the exceptional divisor parametrized by $w$. Notice
that in the double scaling limit $\tau \rightarrow i \infty,
m^{2} \rightarrow \infty $ $\hat{u}$ goes to $\hat{u}=0$ for any
value of $u=\langle Tr \phi^{2} \rangle$. In other words, the
only point in the $(\hat{u},\tau)$-plane of the $N\!=\!2$
theory with $N\!=\!4$ matter content that can have a pure
$N\!=\!2$ interpretation is the enhancement of symmetry
singular point $(\hat{u}=0,\tau=i \infty)$. By means of the blow
up of this singular point we create an extra divisor which
represents the quantum moduli of the pure $N\!=\!2$ theory. 

Secondly, it should be noticed that in $(\epsilon,\hat{u})$
coordinates the curve defined by the locus ${\cal C}_{c}^{(1)}$
and ${\cal C}_{c}^{(2)}$ in the neighbourhood of the point $A$
can be described by $\hat{u}=\pm \epsilon$. This crossing can be
regularized by a single blow up, with the coordinate on the
exceptional divisor given by $w=\frac {\epsilon}{\hat{u}}=\frac
{1}{\tilde{u}}$, and the loci ${\cal C}_{c}^{(1)}$, ${\cal
C}_{c}^{(2)}$ mapped into the lines $w=\pm 1$, which represents
a more natural description of the Seiberg-Witten plane. When we
use $(\epsilon^{2},\hat{u})$ coordinates we effectively quotient
by the $R$-symmetry $\tilde{u} \rightarrow
-\tilde{u}$\footnote{See comment iii) in next section.}
, paying
the price of creating, by the double blow up, the exceptional
divisor (at zero) parametrized by the coordinate $v$. In order
to match the monodromies when we work with
$\tilde{u}^{2}$-coordinates, we need to take into account, as we
move $\tilde{u} \rightarrow e^{2 \pi i} \tilde{u}$ around the
intersection points, the contribution coming from
moving in the $w$-divisor, and the extra piece arising from
moving in the ``orthogonal'' divisor. As we will see in next
section, the reason for considering the blow up in
$(\epsilon^{2},\hat{u})$-coordinates comes from the fact that
the $(\hat{u}, \tau)$-plane defines a double covering of the
Calabi-Yau moduli space. 
  
In the same way as the point $(\hat{u}=0,\tau= i \infty)$ is
used to recover the pure $N\!=\!2$ theory $(f \rightarrow
\infty)$, the line $\{\hat{u}=\infty\}$ can be interpreted as
corresponding to the pure $N\!=\!4$ theory $(f \rightarrow 0)$. 
In this sense, there exits an interesting similarity between the
singular locus ${\cal C}_{0}$ and the ``$N\!=\!4$'' line
$\{\hat{u}=\infty\}$, namely on the line ${\cal C}_{0}$ a component of the elementary hypermultiplet becomes
massless. 


\section{Calabi-Yau Interpretation.}

The interpretation above of the pure $N\!=\!2$ theory as the
blow up of a singular (enhancement of symmetry) point in the
extended moduli of the $N\!=\!2$ theory with $N\!=\!4$ matter 
content, and the results in reference \cite{KKLMV} concerning
the point particle limit, motivate us to compare, in more
detail, the $(\hat{u},\tau)$-plane to the moduli of complex
structures of the Calabi-Yau weighted projective space
$W\IP_{11226}^{12}$. The defining polynomial \cite{can,CYY} is 
\begin{equation}
p=z_{1}^{12}+z_{2}^{12}+z_{3}^{6}+z_{4}^{6}+z_{5}^{2}-12\psi
z_{1}z_{2}z_{3}z_{4}z_{5} - 2\phi z_{1}^{6}z_{2}^{6}.
\end{equation}
Introducing the variables
\begin{equation}
x \equiv \frac {-1}{864} \frac {\phi}{\psi^{6}}, \: \: \: \: \:
\: \: \: y \equiv \frac {1}{\phi^{2}},
\end{equation}
the singular loci are given \cite{can} by\footnote{In
$\phi,\psi$ variables, the four singularity loci of
$W\IP_{11226}^{12}$ are $(1)$ $\{864\psi^6+\phi=\pm1\}$, $(2)$
$\{\phi=\pm1\}$, $(3)$ $\{\phi,\psi=\infty \}$ and $(4)$
$\{\psi=0\}$. In equation (\ref{eq:lomod}) we give these four
loci in the chart with coordinates $(x^{-1},y)$ (see Figure 3).}:
\begin{equation}
        \begin{array}{lcl}
        (1) \: {\cal C}_{con} & \equiv & \{ (1-x)^{2}-x^{2}y=0 \}, \\
        (2) \: {\cal C}_{\infty} & \equiv & \{y=0\}, \\
        (3) \: {\cal C}_{1} & \equiv & \{ y=1\}, \\
        (4) \: {\cal C}_{0} & \equiv & \{ x=\infty \}.
        \end{array}
\label{eq:lomod}
\end{equation}


\begin{figure}[ht]
\def\epsfsize#1#2{.6#1}
\centerline{\epsfbox{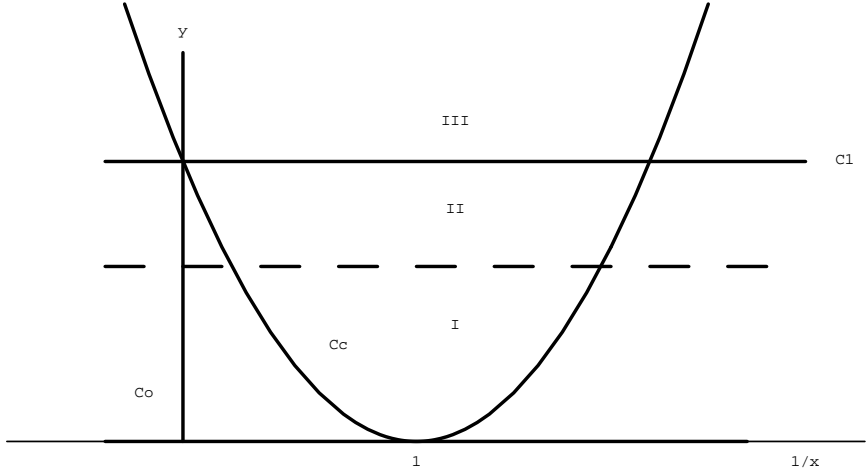}}
\caption{Singular loci of $W\IP^{12}_{11226}$ moduli space.}
\end{figure}


Before entering into a more detailed study of the moduli of
$W\IP_{11226}^{12}$, let us simply consider the blow up of the
tangency point $(x=1,y=0)$ between ${\cal C}_{con}$ and ${\cal
C}_{\infty}$. By means of a two step blow up \cite{KKLMV,can},
we obtain two exceptional divisors $E_{1}$ and $E_{2}$ (see
Figure 4), with coordinates $\frac {y x^{2}}{(1-x)^{2}}$ and
$\frac {yx}{(1-x)}$, respectively. If we are now bold enough to 
identify the blow up in the $(\hat{u},\tau)$-plane (Figure 2)
with the Calabi-Yau blow up of Figure 4, we will get the
following relation\footnote{The variable $\tilde{u}$ entering
the following definition should not be confused with that used
in (\ref{eq:cur}).} 
\begin{equation}
\frac {yx^2}{(1-x)^2} = \frac {\epsilon ^2}{\hat{u}^2} \equiv
\frac {1}{\tilde{u}^2},
\label{eq:44}
\end{equation}
which implies, in the neighbourhoood of the point $(x=1,y=0)$,
\begin{equation}
{\displaystyle x =\frac {1}{1 \pm \hat{u}}+ \cdots} \hspace{1cm}
        y = \epsilon^2 + \cdots
\label{eq:le}
\end{equation}
Notice that relation (\ref{eq:44}) is nothing but the one used in reference
\cite{KKLMV} to define the point particle limit. Here this
relation is derived by simply identifying the blow ups of
Figures 2 and 4. Moreover, using the mirror map of \cite{can},
it was suggested in \cite{KV} the possibility to interpret $y$
in terms of the heterotic dilaton $S$ through $y=e^{-S} +
\cdots$. This identification, together with equation
(\ref{eq:le}), strongly suggests the following string
interpretation of the $N\!=\!2$ theory with $N\!=\!4$ matter content:
\begin{equation}
        \begin{array}{ccc}
        (\alpha')^{-1} & \leftrightarrow & f=\frac {1}{4}m^{2}, \\
        e^{-S}         & \leftrightarrow & 64q=\epsilon^{2}.
        \end{array}
\label{eq:dic}
\end{equation}
Namely, the $N\!=\!4$ bare coupling constant $\tau$ can be
thought of as the dilaton, and the soft breaking mass term $f$
for the hypermultiplets in the adjoint representation as the
inverse of the string tension\footnote{The relation
(\ref{eq:le}) between $x$ and $\hat{u}$ was derived for
$\alpha'=1$ in reference \cite{GL}. The crucial role of
$\alpha'$ in the whole blow up analysis was first pointed out in
reference \cite{KKLMV}.}.


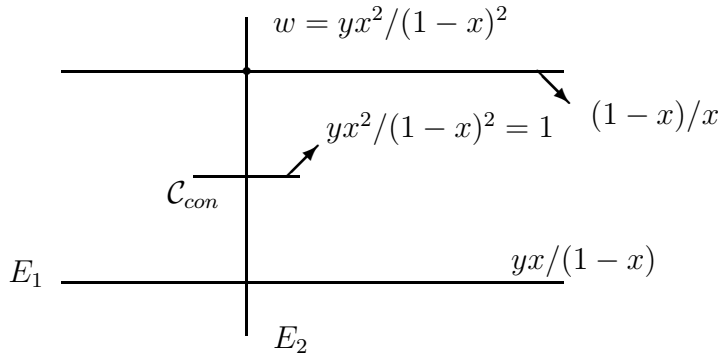
\begin{figure}[htbp]
\centering

     \begin{picture}(400,170)

     \thicklines

     \put(150,20){\line(0,1){120}}  \put(130,80){\line(1,0){40}}
     \put(80,120){\line(1,0){190}}  \put(80,40){\line(1,0){190}}

     \put(165,80){\vector(1,1){12}}  \put(260,120){\vector(1,-1){12}}
     \put(150,120){\circle*{3}}

     \put(160,135){$w=yx^{2}/(1-x)^{2}$}
     \put(280,100){$(1-x)/x$}  
     \put(180,95){$yx^2/(1-x)^2=1$}  \put(250,45){$yx/(1-x)$}
     \put(120,70){${\cal C}_{con}$}
        \put(60,40){$E_{1}$} \put(160,15){$E_{2}$}

     \end{picture}

\caption{The double blow up as seen in the Calabi-Yau variables.}

\end{figure}


The formal similarity between the singular loci in
$(\hat{u},\tau)$-plane and those for $W\IP_{11226}^{12}$ seems
to be more than a coincidence. We can in fact define a one to
one map between the two moduli spaces such that in the weak
coupling limit we recover relations (\ref{eq:le}). This can be
done using the following correspondences:
\begin{equation}
{\displaystyle x=\frac {3/2e_{1}(\tau)}{3/2e_{1}(\tau)-\hat{u}},}
\hspace{1cm} {\displaystyle \sqrt{y}=-
\frac {e_{2}(\tau)-e_{3}(\tau)}{3e_{1} (\tau)},}
\label{eq:ex}
\end{equation}   
which map the loci (\ref{eq:loci}) into the loci
(\ref{eq:lomod}). Some comments are now necessary concerning (\ref{eq:ex}):
\begin{itemize}
        \item[{i)}] To fix the correspondence (\ref{eq:ex}) we
are forced to work with $\sqrt{y}$, i.e., in a double covering.
The reason, already clear in the $\pm$ ambiguity appearing in
(\ref{eq:le}), is that we identify the loci ${\cal C}_{c}^{(1)}$
and ${\cal C}_{c}^{(2)}$ of the $(\hat{u},\tau)$-plane with the
two conifold branches $\sqrt{y}=\pm \left( \frac {1-x}{x}
\right)$. It is important to stress that the ``monopole''
and ``dyon'' singularities, in the $N\!=\!2$ field theory
language, are identified in the $(x,y)$-plane.
        \item[{ii)}] In the coordinate chart II $(x,x^2y)$ of
the corresponding toric diagram \cite{can}, the strong coupling
loci ${\cal C}_{1}$ is tangent to ${\cal C}_{\infty}$ at the
origin. This tangency can be again blown up in two steps. We
get, in this way, the extra divisor $\{x=0\}$. By (\ref{eq:ex})
this divisor corresponds to the loci $\{\hat{u}=\infty\}$ in the
$(\hat{u},\tau)$-plane. Recalling now that $\hat{u}$ was defined
as $u/f$, this loci corresponds to the pure $N\!=\!4$ limit $f
\rightarrow 0$.
 
The two step blow up in the coordinate chart II produces, as
mentioned above, two exceptional divisors, $\{x=0\}$ and
$D_{(-1,-1)}$ (in the notation of reference \cite{can}). This
second divisor has no analog in the $(\hat{u},\tau)$-plane. The
reason is again that the $(\hat{u},\tau)$-plane is properly
speaking in correspondence with the plane $(x,\sqrt{y})$, where
the tangency we are working out becomes a crossing with a single
step blow up, and one exceptional divisor, which is the divisor
$\{x=0\}$. As mentioned above, the $(\hat{u},\tau)$-plane
defines a double covering; thus, the second divisor
$D_{(-1,-1)}$ in the Calabi-Yau moduli will only appear when we
quotient by the covering transformation. 
        \item[{iii)}] By correspondence (\ref{eq:ex}), the
$(\hat{u},\tau)$-plane becomes a double covering of the
$(x,y)$-plane. The covering transformation is precisely given by
the $T$ element in the $Sl(2,Z)$ duality group,
$T:(\hat{u},\tau) \rightarrow (\hat{u},\tau+1)$. Moreover, the
action of $T$ corresponds to the map $A:(\phi,\psi) \rightarrow
(- \phi,\alpha \psi)$, with $\alpha^{12}=1$, i.e., to the transformation between the two
branches $\sqrt{y}=\pm \left( \frac {1-x}{x} \right)$ of the
conifold locus. This explains why
the point particle limit defined as the blow up of the tangency
point produces a quantum moduli parametrized in terms of
$\tilde{u}^2$ instead of $\tilde{u}$. More precisely the parametrization
 in terms of $\tilde{u}^2$ does not mean that we can quotient in
the rigid theory by the global $R$-symmetry $u\rightarrow -u$.
To arrive to $\tilde{u}^2$ we need first to go to the extended
moduli $(\hat{u},\tau)$, secondly to notice that the $(\hat{u},\tau)$-plane
is a double covering of the $(x,y)$- moduli space and third to
quotient by the ``stringy'' symmetry $A$ which corresponds to
the covering map $T:(\hat{u},\tau) \rightarrow (\hat{u},\tau+1)$.
        \item[{iv)}] Taking into account the transformation
rules of the roots $e_{i}$ we can use the correspondences
(\ref{eq:ex}) to map different regions of $H^{+}/\Gamma_{2}$
(see Figure 5) into the $(x,y)$ moduli space. So, the domain
I:$[\tau=i,\tau=i \infty]$ goes into the domain $[y=0,y=1/3]$.
In just the same way, the interval $[i,0]$ goes into the domain
$[y=1/3,y=1]$. For the domain III of Figure 5, we consider the
line going from $\tau=0$ to the point $\tau=\frac {1}{2}(i-1)$;
this point is the $ST$ transformed of $\tau=i$. Using
(\ref{eq:ex}) we see that $y(\tau=\frac {1}{2}(i-1))=\infty$,
and therefore we map the domain III of Figure 5 into the region
$[y=1,\infty]$.

\begin{figure}[ht]
\def\epsfsize#1#2{.7#1}
\centerline{\epsfbox{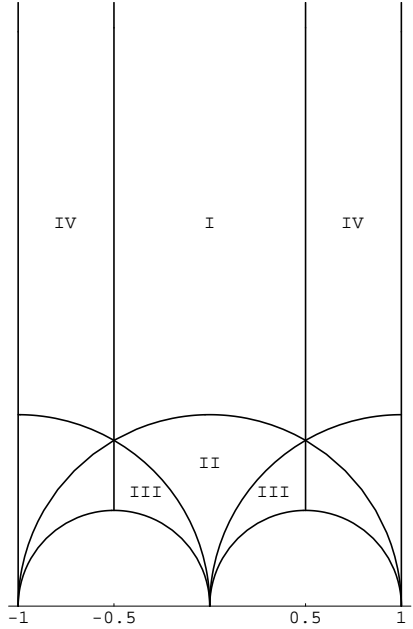}}
\caption{Modular domain of $\Gamma_2$.}
\end{figure}

        \item[{v)}] The previous discussion can be repeated
``mutatis mutandis'' for the Calabi-Yau weighted projective
space $W\IP_{11222}^{8}$. The relevant difference between these
two spaces is the modular group in the limit $y\rightarrow 0$,
which is $Sl(2,Z)$ for $W\IP_{11226}^{12}$, and
$\Gamma_{0}(2)_{+}$ for $W\IP_{11222}^{8}$ \cite{KLM}. In both
cases we can recover the rigid $SU(2)$ Seiberg-Witten solution
\cite{KKLMV,AP}. The enhancement of symmetry point for
$W\IP_{11226}^{12}$ is given by $T_{heterotic}=i$, while for
$W\IP_{11222}^{8}$ is $T_{heterotic}=i/\sqrt{2}$, reflecting the
difference in the mirror map: Jacobi's j function for
$W\IP_{11226}^{12}$, and the Haupmodul for $\Gamma_{0}(2)_{+}$
in the case of $W\IP_{11222}^{8}$. In the above construction,
leading to the correspondence (\ref{eq:ex}), these differences
between $W\IP_{11226}^{12}$ and $W\IP_{11222}^{8}$ are not taken
into account; we will come back to this point in next section.
        \item[{vi)}] As a last comment on the correspondence
(\ref{eq:ex}), we consider the behaviour at $\tau=\frac
{1}{2}(i-1)$. Since $e_{1}(\frac {1}{2}(1-i))=0$, the point
$(\hat{u}=0,\tau =\frac {1}{2}(1-i))$, which is in the locus
${\cal C}_{0}$, blows up by (\ref{eq:ex}) into the whole line
$(x,y=\infty)$. This fact has indeed its counterpart in the
Calabi-Yau moduli space. Namely, at the point $y=\infty$
$(\phi=0)$, the locus ${\cal C}_{0}=\{\psi = 0\}$ corresponds to
an undetermined value of $x=-\frac {\phi}{864\psi^6}$.

\end{itemize}


\section{$S$-Duality and Yukawa Couplings.}

We will use the map (\ref{eq:ex}) to induce the action of the
$S$-duality group $Sl(2,Z)$, acting on the $N\!=\!4$ moduli
$\tau$, on the Calabi-Yau space.
  
As already mentioned in the previous paragraph, points related
by the $T$-transformation $(\hat{u},\tau) \rightarrow
(\hat{u},\tau+1)$ map into the same $(x,y)$ point. The $T$
action is in fact non trivial only when we work in the double
covering space with coordinates $(x,\sqrt{y})$. On this space
the $T$ transformation is interchanging the two branches of the
conifold locus, and the two branches $\sqrt{y}=\pm 1$ of the
singular locus ${\cal C}_{1}$.

The non perturbative generator $S:\tau
\rightarrow -\frac {1}{\tau}$ induces the change
\begin{equation}
   \begin{array}{ccl}
   x(\hat{u},\tau) & \rightarrow & x(\hat{u}^M,-\frac{1}{\tau})
   \equiv x'(\hat{u},\tau), \\
   y(\tau) & \rightarrow & y(-\frac{1}{\tau}) \equiv y'(\tau).
   \end{array}
\label{SD}
\end{equation}
where
\begin{equation}
\hat{u}^M=\tau ^{2} \left(\hat{u}+\frac {1}{2}(e_{2}(\tau) 
-e_{1}(\tau))\right)
\end{equation}
has been defined using the transformation of $\tilde{u}$
$(u=\tilde{u}+ \frac {1}{2}e_{1}f)$ as a modular form of weight
two. From the map (\ref{eq:ex}) it is now easy to derive
\begin{equation}
{\displaystyle x'=\frac {1}{2} \frac {1+3\sqrt{y}}{1+\sqrt{y}-\frac
{1}{x}},} \hspace{1cm}
{\displaystyle \sqrt{y}'= \frac {1- \sqrt{y}}{1+3\sqrt{y}}.}
\label{eq:xytr}
\end{equation}

The transformations of the different loci under $S$, as defined
by (\ref{eq:xytr}), are given as follows. The positive branch
$\sqrt{y}= + \left( \frac {1-x}{x} \right) $ of the conifold
locus is mapped by $S$ into ${\cal C}_{0}$, while the negative
branch $\sqrt{y}= - \left( \frac {1-x}{x} \right) $ is mapped
into itself. In a similar way, the negative branch $\sqrt{y}=-1$
of the locus ${\cal C}_{1}$ is mapped into itself, while the
positive branch $\sqrt{y}=+1$ is mapped into ${\cal
C}_{\infty}$. Moreover, the enhancement of symmetry point
$(x=1,y=0)$ is mapped by $S$ into the point of crossing between
${\cal C}_{0}$, ${\cal C}_{1}$ and ${\cal C}_{con}$.
  
Now we pass to study the action of the $S$ duality group
$Sl(2,Z)$ on some geometrical objects. In what follows, we will
reduce ourselves to the Yukawa couplings; this check will again
be unable to distinguish between $W\IP_{11226}^{12}$ and
$W\IP_{11222}^{8}$, since the couplings of both spaces are
identical up to a global factor of four \cite{can,CYY}. From reference
\cite{CYY} we take\footnote{These couplings are obviously
invariant under the $T$ transformation.}
\begin{equation}
Y_{xxx} = {\displaystyle \frac {1}{x^{3}\Delta},} \hspace{1cm}
Y_{xxy} = {\displaystyle \frac {1-x}{2x^{2}y \Delta},} \hspace{1cm}
Y_{xyy} = {\displaystyle \frac {2x-1}{4xy(1-y) \Delta}.}
\label{eq:YC}
\end{equation}
where $\Delta\!=\!(1-x)^2 - x^2 y$ is the conifold locus
discriminant. Under the $S$-transformations (\ref{eq:xytr})
the Yukawa couplings are transformed as follows:
\begin{equation}
        \begin{array}{c}
        Y_{x'x'x'}=\left( \frac {\partial x}{\partial x'}
        \right) ^{3} Y_{xxx} = \frac {1}{x'^{3} \Delta'}, \\
        Y_{x'x'y'} = \left( \frac {\partial x}{\partial x'}
        \right) ^{2} \left( \frac {\partial x}{\partial y'}
        \right) Y_{xxx} + \left( \frac {\partial x}{\partial
        x'} \right)^{2} \left( \frac {\partial y}{\partial y'}
        \right) Y_{xxy} = \frac {1-x'}{2x'^{2}y' \Delta'}
        \left( \frac {- \sqrt{y}'}
        {(1-x')(1-\sqrt{y}')} \right), \\
        Y_{x'y'y'}= \left( \frac {\partial x}{\partial x'} \right)
        \left( \frac {\partial x}{\partial y'} \right)^2 Y_{xxx}
        + 2 \left( \frac {\partial x}{\partial x'} \right)
        \left( \frac {\partial x}{\partial y'} \right)
        \left( \frac {\partial y}{\partial y'} \right) Y_{xxy} +
        \left( \frac {\partial x}{\partial x'} \right) \left(
        \frac {\partial y}{\partial y'} \right)^2 Y_{xyy} = \\
        = \frac {2x'-1}{4x'y'(1-y')} \frac{1}{\Delta'}
        \left( \frac{2x'-1-(1+\sqrt{y'})}{(2x'-1)\sqrt{y'}} \right)
        \end{array}
\label{eq:YCtr}
\end{equation}

In the above calculations we have made use of the following property,
derived from the transformations (\ref{SD}):
\begin{equation}
\Delta(x',y') = \frac{\Delta(x,y)}{(1-x(1+\sqrt{y}))^3}
\label{1}
\end{equation}
As an example, using $(\frac{dx}{dx'})\!=\!(1-x(1+\sqrt{y}))x/x'$
and (\ref{1}), it is immediate to obtain $Y_{x'x'x'}$ as given in
(\ref{eq:YCtr}).

{}From (\ref{eq:YC}) and (\ref{eq:YCtr}) we observe that $Y_{xxx}$
is invariant, while the others pick up, by the $S$-duality
transformation, an extra factor. It is important to observe the
non trivial fact that in all cases the extra factor becomes
{\em one on the negative branch of the conifold locus\/}. This
branch is precisely the locus which is mapped into itself by the
action of the $S$-duality transformation. In this sense, we can
interpret results (\ref{eq:YCtr}) as reflecting the different
``modular weights'' of the Yukawa couplings with respect to the
$S$-duality group $Sl(2,Z)$ \cite{wp}, a difference that should
certainly vanish on the locus that is mapped into itself by the
$S$-duality transformation. Moreover, it should be noticed that
the extra factors in (\ref{eq:YCtr}) are due to the fact that
the transformations (\ref{eq:xytr}) are defined on the double
covering. 


\section{Comments.}

To conclude this letter we reduce ourselves to mentioning some
aspects of our analysis that deserve a deeper understanding.
        \begin{itemize}
        \item[{i)}] The physical picture, as described in the
$(x,y)$ plane, differs from the one we will obtain in its double
cover in many aspects. In particular, the $S$-duality action, as
defined in this letter, can only be implemented on the double
covering; namely the $T$ part of the $Sl(2,Z)$ duality group is
precisely the transformation interchanging the two branches of
the double covering. A second aspect related with the double
covering goes to the more technical point on the blow up of the
tangency between ${\cal C}_{1}$ and ${\cal C}_{\infty}$. If we
work in the double covering, we only need one exceptional
divisor, which is precisely the one describing the pure
$N\!=\!4$ theory.
        \item[{ii)}] It would also be interesting to work out
the physics of the singular locus ${\cal C}_{0}$, which in our
approach is the $S$-dual of the negative branch of the conifold,
a fact again hidden by working in $(x,y)$ variables.
        \item[{iii)}] The $N\!=\!4$ to $N\!=\!2$ flow framework we
        have used in our study is, as was pointed out in
        reference \cite{DW}, intimately connected with the
        relation between integrable models and $N\!=\!2$
        theories \cite{int,int2}. A natural question that appears
from our study
will be the connection between these integrable models and the
Calabi-Yau manifold used to reinterpret the $N\!=\!4$ to
$N\!=\!2$ flow. Moreover, it would be important to understand
whether the connection between $N\!=\!2$ gauge theories and
integrable models can be reinterpreted from the point of view of
the underlying string theory.
       
        \item[{iv)}] At a more speculative level, we can wonder
whether the stringy interpretation of the $N\!=\!4$ to $N\!=\!2$
flow described in this letter can be extended to a flow from
$N\!=\!2$ to $N\!=\!0$ or $N\!=\!1$.
        \item[{v)}] The role of $N\!=\!4$ in our approach, and
in this sense the physics of the double covering, can be perhaps
understood if we think of the $N\!=\!4$ theory as the
dimensional reduction of $N\!=\!2$ in $D=5$, in the spirit of
reference \cite{r2}. The strong coupling spectrum is 
interpreted ,in Kaluza-Klein terms, in parallel to the $D=11$
interpretation of the strong coupling regime in string theory.
It would also be interesting to have a $D$-brane interpretation
\cite{POL,PW} of the hypermultiplets in the adjoint that control
the strong coupling loci in a similar way to Strominger's
interpretation \cite{bh} of the conifold locus.
        \end{itemize}

\vspace{15 mm}

This work is partially supported under grant by European
Community grant ERBCHRXCT920069, by PB92-1092. The work of E. L.
is supported by M. E. C. fellowship. The work of R. H. is
supported by U.A.M. fellowship.

\newpage


\end{document}